\documentclass[twocolumn,amsmath,amssymb,showpacs]{revtex4}
\usepackage{graphicx}% Include figure files
\usepackage{dcolumn}% Align table columns on decimal point
\usepackage{bm}% bold math
\begin{document}

\title{Metal nanofilm in strong ultrafast optical fields}

\author{Vadym Apalkov}
\author{Mark I. Stockman}
\affiliation{
Department of Physics and Astronomy, Georgia State
University, Atlanta, Georgia 30303, USA
}

\date{\today}
\begin{abstract}
We predict that a metal nanofilm subjected to an ultrashort (single oscillation) optical pulse of a high field amplitude $\sim 3~\mathrm{V/\AA}$ at normal incidence undergoes an ultrafast (at subcycle times $\lesssim 1~\mathrm{fs}$) transition to a state resembling semimetal. Its reflectivity is greatly reduced, while the transmissivity and the optical field inside the metal are greatly increased. 
%The deposition of the pulse energy as a function of field reaches its maximum and then decreases due to suppression of the active optical resistance. 
The temporal profiles of the optical fields are predicted to exhibit pronounced subcycle oscillations, which are attributed to the Bloch oscillations and formation of the Wannier-Stark ladder of electronic states. The reflected, transmitted, and inside-the-metal pulses have non-zero areas approaching half-cycle pulses. The effects predicted are promising for applications to nanoplasmonic modulators and field-effect transistors with petahertz bandwidth.
\end{abstract}

\maketitle 

Behavior of solids in strong ultrafast optical fields has recently attracted a great deal of attention  \cite{Krausz_et_al_PRL_1998_Femtosecond_Breakdown_Dielectrics, Murnane_et_al_PRL_97_113604_2006_Laser_Assisted_Photoelectric_Effect_from_Surfaces, Corkum_et_all_JOPB_Attosecond_Ionization_SiO2, Baltuska_et_al_Attosecond_IonizationPRL_2011, Reis_et_al_Nature_Phys_2011_HHG_from_ZnO_Crystal, Stockman_et_al_PRL_2010_Metallization, Hommelhoff_et_al_Nature_2011_As_Tip_Electron_Emission, Kling_et_al_Nature_Phot_2011_Dielectric_Sphere_Ultrafast_Photemission, Stockman_et_al_PRL_2011_Dynamic_Metallization, Schiffrin_at_al_Nature_2012_Current_in_Dielectric, Schultze_et_al_Nature_2012_Controlling_Dielectrics}. Such fields produce non-perturbative effects on solids, among which are ultrafast optical breakdown \cite{Krausz_et_al_PRL_1998_Femtosecond_Breakdown_Dielectrics}, attosecond ionization \cite{Corkum_et_all_JOPB_Attosecond_Ionization_SiO2, Baltuska_et_al_Attosecond_IonizationPRL_2011}, metallization of dielectric nanofilms \cite{Stockman_et_al_PRL_2010_Metallization, Stockman_et_al_PRL_2011_Dynamic_Metallization},  optical field-effect reversible  subfemtosecond currents in dielectrics \cite{Schiffrin_at_al_Nature_2012_Current_in_Dielectric}, and electron tunneling from surfaces \cite{Murnane_et_al_PRL_97_113604_2006_Laser_Assisted_Photoelectric_Effect_from_Surfaces, Hommelhoff_et_al_Nature_2011_As_Tip_Electron_Emission, Kling_et_al_Nature_Phot_2011_Dielectric_Sphere_Ultrafast_Photemission}. For dielectrics, when optical field is applied with frequency $\hbar\omega$ low enough compared with the band gap $\Delta_{vc}$ between the valence a conduction bands, mostly adiabatic processes take place such as Wannier-Stark (WS) localization and formation of the WS ladder of levels \cite{Wannier_1959_Book_Solid_State, Wannier_PR_1960_Wannier_States_in_Strong_Fields} separated by the Bloch frequency \cite{Bloch_Z_Phys_1929_Functions_Oscillations_in_Crystals} $\omega_B= |e|Fa/\hbar$, where $e$ is electron charge, $F$ is the magnitude of the field, and $a$ is the lattice constant. Only when the field $F$ exceeds critical field $F_c=\Delta_{vc}/(|e|a)\sim 2.5~ \mathrm{V/\AA}$ (for $\Delta\sim 10$ eV and $a\sim 4~\mathrm{\AA}$),  the band gap is overcome by the WS splitting, and the diabatic coupling of the valence and conduction band becomes strong, which can lead, in particular, to optical breakdown  \cite{Schiffrin_at_al_Nature_2012_Current_in_Dielectric}. 

In contrast, this Letter deals with strong optical fields applied to good (plasmonic) metals where there is no band gap at the Fermi surface and, consequently, no adiabaticity for relatively low fields. In such a case, there are a high optical conductivity and a skin layer  with a depth $l_s\sim 25$ nm \cite{Stockman_Opt_Expres_2011_Nanoplasmonics_Review}. Consequently for metal thickness $h\gtrsim l_s$, a significant or dominating fraction of the incident radiation energy is reflected. Interaction of the radiation with the metal becomes adiabatic only when the optical field is strong enough so $\omega_B\gg \omega$. The plasmonic metal behavior seizes and WS localization \cite{Wannier_1959_Book_Solid_State, Wannier_PR_1960_Wannier_States_in_Strong_Fields} is established when, during a quarter optical period $t=\pi/(2\omega)$, an electron acquires momentum $\pi |e| F t$ that exceeds the width $2\pi\hbar/a$ of the Brillouin zone. This condition is satisfied when the optical field $F$ exceeds a critical field  $F_c=4\hbar\omega/(|e|a)\sim 2~\mathrm{V/\AA}$ for  $\hbar\omega=1.55$ eV. Note that at this field the WS states are already strongly localized, $l_{WS}\ll a$, where $l_{WS}=\hbar^2/\left(m a^2 |e|F\right)$ is the WS localization radius, and $m$ is electron mass. For $F\gtrsim F_c$, the strong-field regime for the metal sets on. 

As we predict in this Letter, in the strong-field regime the optical properties of the metal differ dramatically from those at low to moderate fields, becoming reminiscent of semimetals: plasmonic properties and strong reflection associated with the skin effect are suppressed during subcycle time intervals driven by the {\em instantaneous} optical field.  Light transmission through the metal is increased but the  optical absorption in the metal is reduced at very high fields. Ultrafast behavior of the metal is radically changed: both the reflection and transmission exhibit subcycle Bloch-type oscillations with period $\tau_F\sim 2\pi/\omega_B$; e.g., $\tau_F\sim 0.5$ fs for $F=2.5~\mathrm{V/\AA}$. Thus, the strong optical field is predicted to cause ultrafast response of the metal irrespectively of the frequency and duration of the excitation pulse. Under these conditions, the Bloch oscillations are predicted to manifest themselves in natural metals, while earlier such oscillations  were observed only in artificial semiconductor superlattices  -- see, e.g., Refs.\ \onlinecite{Mendez_Bastard_Phys_Today_1993_WS_Ladders_in_Superlattices, Feldmann_et_al_PRB_1992_Bloch_Oscillations_Observation_in_Superlattices, Dekorsy_et_al_SST_1994_Boch_Oscillations}. These predicted strong-field effects open up routes toward using metals as active elements for deep ultrafast modulation of optical fields -- cf. the perturbative modulation in active plasmonics \cite{MacDonald2009_et_al_Nature_Photonics_3_55_2009_Ultrafast_Active_Plasmonics}.

Consider an ultrashort optical pulse incident normally on a metal nanofilm. Propagation 
of such a pulse is described by the Maxwell equations where dielectric polarization is determined by quantum dynamics of the electrons in the metal. This dynamics is described by the  Schr\"odinger equation in the presence of the electric field inside the metal. We neglect the Coulomb scattering of electrons because the characteristic time $\tau_s$ of such scattering in metals significantly exceeds the length of our optical pulse (e.g., $\tau_s=20-50~\mathrm{fs}$ in silver \cite{Stockman_Opt_Expres_2011_Nanoplasmonics_Review}).

We solved numerically the coupled system of the Maxwell and Schr\"odinger equations using the finite difference time domain (FDTD) method \cite{Kunz_Luebbers_1993_FDTD, Taflove_FDTD_Book} for a finite-size system with the
absorbing boundary conditions for Maxwell equations. The tight-binding model was used in the solution of the Scr\"odinger equation. The size of the computational space in the direction of propagation of the pulse
($z$ direction) was 6000 nm. 
The metal film was placed at the midplane of the system, i.e., at $z = 0$. 
In numerical solution of the Maxwell equations, we assumed that the spatial step was 1 nm and the time step was 0.7 attoseconds (1 as=$10^{-18}$ s). 
The optical pulse was generated at the left boundary and propagated along the positive 
direction of the $z$ axis with the polarization of the electric field along the $x$ axis. 

We assume that a single-oscillation pulse form,  
\begin{equation}
F_{x}(t) = F_{0} e^{-u^2} \left( 1 - 2 u^2 \right),
\label{FV0}
\end{equation}
where $F_0$ is the amplitude, 
which is related to power ${\cal P} =c F_0^2/4 \pi$, where $c$ is speed of light,
$u = t/\tau $, and $\tau $ is the pulse length, which is set $\tau = 1 $ fs in our  calculations. Similarly short pulses have been recently used \cite{Schiffrin_at_al_Nature_2012_Current_in_Dielectric}.
%Due to the ultrashort duration of this pulse,  the electron scattering, 
%which is determined by much longer characteristic time, $\tau_e\approx 20$ fs -- see, e.g., Ref.\ %\onlinecite{Stockman_Opt_Expres_2011_Nanoplasmonics_Review}, can be neglected. In this case, 
The metal can be described by one-particle Schr\"odinger equation with 
the Hamiltonian 
\begin{equation}
{\cal H} = \frac{ {\bf p}^2}{2m} + V({\mathbf r}) + e F_x (z,t) x~,
\label{H1}
\end{equation}
where $V(\mathbf{r})$ is the periodic crystal potential, and $F_x (z,t)$ is the electric field inside the metal, which is found from the solution of the Maxwell equations.
Without the electric field, the electron system has standard band structure. Below we consider one conduction band (CB, or sp-band in silver) and one valence band (VB, or d-band).
% The electric field, $F_x(z,t)$, introduces coupling between different electron states of the bands. 

We assume that the periodic potential $V(\mathbf{r})$ is separable in all three directions, $x$, $y$, and $z$,  with period $a$. Then the electron dynamics along the direction $x$ of electric field separates. For each band, the energy dispersion law has the tight-binding 
form \cite{Slater_Koster_PR_1954_LCAO_or_Tight_Binding, Frauenheim_et_al_PSS_2000_Tight_Binding_Review}
$%\begin{equation}
E_{\alpha }(k) =\epsilon _{\alpha } + \frac{\Delta_{\alpha }}{2} \cos (ka)$,
%\end{equation}
where $\alpha = c$ or $v$  for CB and VB, respectively, $\Delta_{\alpha }$ is the width of band $\alpha$,  and 
$\epsilon _{\alpha }$ is the band offset. In the absence of the optical field, the wave functions satisfy the Bloch theorem, 
%\begin{equation}
$\psi_{\alpha k} (x) = \frac{1}{2\pi} e^{i k x} u_{\alpha k}(x)~$, 
%\label{psiBloch}
%\end{equation}
where $u_{\alpha k}(x+a) =u_{\alpha k} (z) $ are periodic Bloch unit-cell functions,  and $k$ is the (pseudo) wave vector.

In the presence of the optical field, $F_x (z,t)$, we solve numerically the 
time-dependent Schr\"odinger equation by using  the Houston functions \cite{Houston_PR_1940_Electron_Acceleration_in_Lattice} as the basis and describing the 
coupling of VB and CB in terms of the dipole matrix elements $Z_{\alpha\alpha ^{\prime}}$ -- 
see Eqs.\ (1)-(5) of Appendix. With the known time-dependent wave functions, we compute polarization and current in a standard way to substitute into for Maxwell equations (see Appendix), thus closing the problem.

Below we use model parameters corresponding to the 
band structure of silver: 
$\epsilon _{v} = -4.1$ eV, $\epsilon _{c} = 0$ eV, $\Delta_{v } = 0.82$ eV, and 
$\Delta _{c} = -9.1$ eV. 
We choose $Z_{vc} = 0.7 ~e  \mathrm{\protect{\AA}}$  \cite{Cooper_at_al_PRB_1965}.
The thickness of the film is set $h=25$ nm. 

%------------------------------------------------------
\begin{figure}
\begin{center}\includegraphics[width=0.4\textwidth]{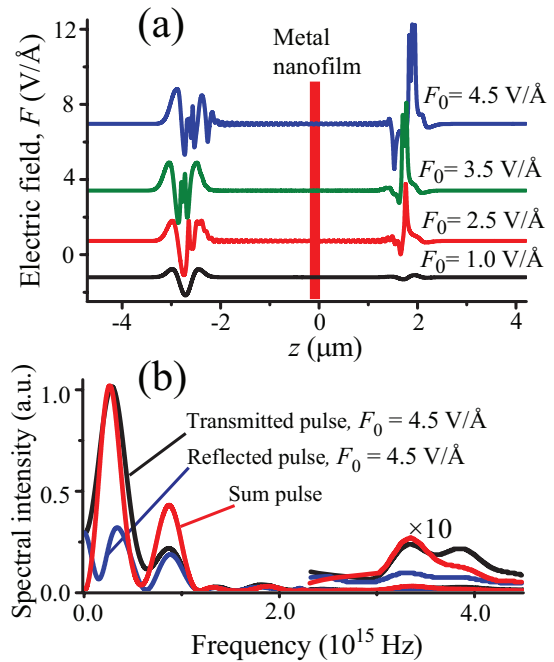}\end{center}
%\vspace*{-1cm}
\caption{Reflected and transmitted pulses. (a)
Spatial distributions of the electric field as functions of the propagation coordinate $z$ shown for different values of $F_0$. The 
metal film of thickness 25 nm is placed at the center ($z=0$) and depicted as the red stripe. The distribution of electric field consists of the reflected (to the left) and transmitted (to the right) pulses propagating in the opposite directions. 
The size of the computational field in the $z$ direction is 6000 nm. (b) Spectral intensities of reflected, transmitted, and sum pulses for $F_0=4.5~\mathrm{V/\AA}$ as functions of optical frequency $f=\omega/(2\pi)$. The transmitted- and sum-pulse spectra are arbitrarily normalized to unity maximum. The reflected pulse spectrum is normalized by the same coefficient as the transmitted one. 
Fragments of the curves at $f>2~\mathrm{PHz}$ are also shown with the $\times10$ magnification as indicated on the graph.
} 
\label{Fig_field_distribution_different_F}
\end{figure}
%----------------------
 
In Fig.\ \ref{Fig_field_distribution_different_F} (a), the spatial distribution of the pulse electric field is shown for the reflected (left) and transmitted (right) pulses for amplitudes $F_0$. For a relatively small field, $F_0=1~\mathrm{V/\AA}<F_c$, the nanofilm behaves as a regular 
metal with a pronounced skin effect and strong reflection of the incident pulse. With increasing the field, $F_0=2.5~\mathrm{V/\AA}\gtrsim F_c$, the response of the electron system is highly nonlinear, and the metal film becomes relatively transparent. Both the reflected and transmitted pulses are strongly reshaped compared to the incident pulse.  As the pulse peak field further increased to $F_0=3.5~\mathrm{V/\AA}$ and $F_0=4.5~\mathrm{V/\AA}$, the film transparency is further increased, in a sharp contrast to metallic behavior. Also, there are pronounced sub-wavelength oscillations in the pulse shape for both the reflected and transmitted fields, which we interpret as an effect of the Bloch oscillations. 

Importantly,  in Fig.\ \ref{Fig_field_distribution_different_F} (a) there are nonzero areas of each transmitted and reflected pulses (denoted by $t$ and $r$, correspondingly), $\Theta^{(t,r)}=\int_{-\infty}^\infty F(z^{(t,r)},t)dt\ne 0$, where $z^{(t)}>0$ and $z^{(r)}<0$. This is due to nonlinearity of the field interaction with the metal. Note that the area of the incident laser pulse of Eq.\ (\ref{FV0}) is exactly zero.  When the absorption in the matter is small, which is the case presently, then $\left|\Theta^{(r)}+\Theta^{(t)}\right|\ll \left|\Theta^{(r)}\right|+\left|\Theta^{(t)}\right|$. This means that the nonlinear transmission and reflection of the metal nanofilm separates the zero-area laser pulse into two pulses (transmitted and reflected) with the non-zero and approximately opposite areas.

Pulses with $\Theta\ne 0$ do not contradict Maxwell equations, and they fundamentally can exist. For instance, for  an optically-linear uniform medium, a plane wave with fields $E_x=H_y=f(z-tc)$, where $f$ is an arbitrary function, and $c$ is speed of light, is a general solution of Maxwell equations. 
Experimentally, near unipolar, half-cycle electromagnetic pulses were generated by aperiodic acceleration of electrons in photo-conductive switches in terahertz spectral region \cite{Bucksbaum_et_al_PRL_1993_Half-Cycle_Pulses,
Raman_et_al_PRL_1996_Half_cycle_pulse_ionization_non_zero_area_exp}. Such pulses accelerate and transfer momentum and energy to free and quasi-free electrons such as those in Rydberg states \cite{Bucksbaum_et_al_PRL_1993_Half-Cycle_Pulses,
Raman_et_al_PRL_1996_Half_cycle_pulse_ionization_non_zero_area_exp}. 

 The magnitudes and signs of the predominant fields for both transmission and reflection are determined by carrier-envelope phase $\varphi_{CE}$ of the excitation pulse, as characteristic for nonlinear effects in a few-oscillation fields, cf. Refs.\ \onlinecite{Schiffrin_at_al_Nature_2012_Current_in_Dielectric, Schultze_et_al_Nature_2012_Controlling_Dielectrics}. Our laser-source pulses possess $\varphi_{CE}=0$, see Eq.\ (\ref{FV0}), and nonlinearity is such that both absorbance and reflectance decrease with the field (cf.\   Fig.\ \ref{Fig_reflectance} below); consequently, $\Theta^{(t)}>0$ and $\Theta^{(r)}<0$. For  $\varphi_{CE}=\pi$, the sign of the dominant field would change to the opposite yielding  $\Theta^{(t)}<0$ and $\Theta^{(r)}>0$. For  $\varphi_{CE}=\pi/2$, both the transmitted and reflected pulses have zero areas. The present effect can be used to generate near-half-cycle pulses in near-infrared and visible.

In Fig.\ \ref{Fig_field_distribution_different_F} (b), we display spectral intensities of the transmitted and reflected pulses $I^{(t,r)}(f)=\left |F^{(t,r)}_f\right|^2$, where $f=\omega/(2\pi)$ indicates Fourier transform in terms of linear frequency.
Note that at $f=0$, $I^{(t,r)}(0)=\left[\Theta^{(t,r)}\right]^2\ne 0$; this confirms the nonzero areas of the transmitted and reflected pulses discussed above in the previous paragraph.

Besides peaks at the carrier frequency, $f\approx 0.25$ PHz,  in Fig.\ \ref{Fig_field_distribution_different_F} (b) there are  peaks at  the approximately the third harmonic frequency, $f\approx 0.75$ PHz, which are due to the strong nonlinearity. There are also smaller but still appreciable peaks at the Bloch frequency, $f\approx 3.5~\mathrm{PHz}$, which are shown magnified by a factor of $\times 10$. Observation of these peaks, which stem from the Bloch oscillations \cite{Bloch_Z_Phys_1929_Functions_Oscillations_in_Crystals, Mendez_Bastard_Phys_Today_1993_WS_Ladders_in_Superlattices, Feldmann_et_al_PRB_1992_Bloch_Oscillations_Observation_in_Superlattices, Dekorsy_et_al_SST_1994_Boch_Oscillations}, would be the first evidence of the Bloch oscillations in real crystals. 

%------------------------------------------------------
\begin{figure}
\begin{center}\includegraphics[width=0.4\textwidth]{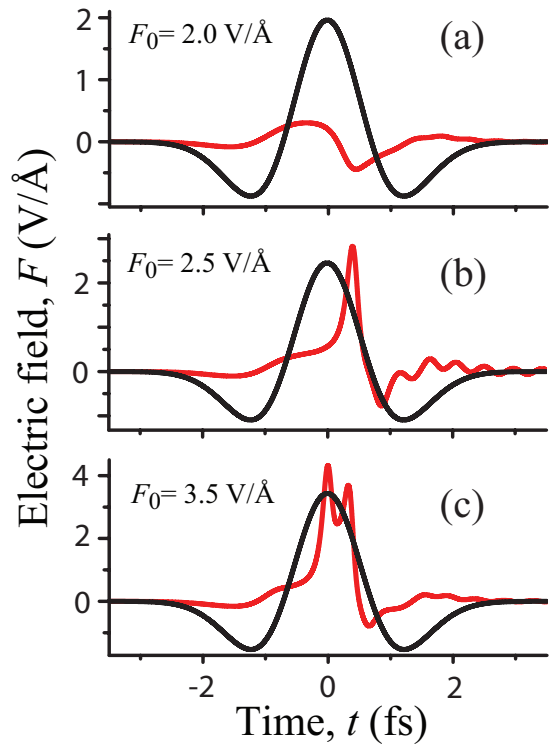}\end{center}
%\vspace*{-1cm}
\caption{The electric field of the incident pulse (black lines) and the electric field at the midpoint of the metal 
film (red line) are shown for different values of the peak electric field $F_0$ of the incident  
pulse: (a) $F_0 = 2.0$ $V/\mathrm{\protect{\AA}}$, (b) $F_0 = 2.5$ $V/\mathrm{\protect{\AA}}$, and 
(c) $F_0 = 3.5$ $V/\mathrm{\protect{\AA}}$.
Note that the pulse areas are non-zero, cf.\ discussion of Fig.\ \ref{Fig_field_distribution_different_F}, which causes net current and charge transfer along the metal in the $x$ direction.
}
\label{Fig_field_inside_outside}
\end{figure}
%----------------------

%------------------------------------------------------
\begin{figure}
\begin{center}\includegraphics[width=0.4\textwidth]{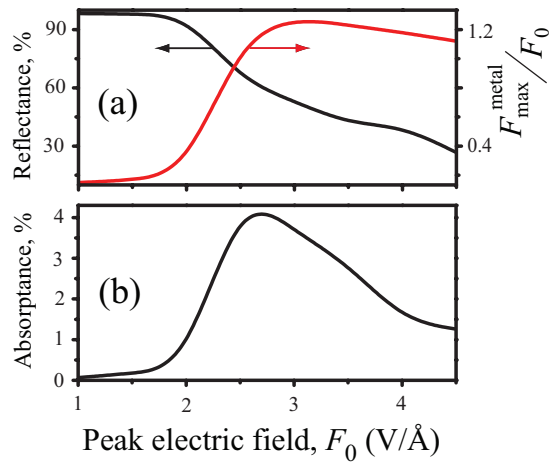}\end{center}
%\vspace*{-1cm}
\caption{
(a) The reflectance of optical pulse (black line) and the maximum electric field at the midpoint of the nanofilm 
(red line) are shown as functions of the peak electric field $F_0$ of the incident  
pulse. The maximum electric field in the metal film is shown in units of the peak electric field $F_0$. 
%The thickness of the metal film is 25 nm. 
(b) The absorbance of the optical pulse is shown as a function of the peak electric field $F_0$.
}
\label{Fig_reflectance}
\end{figure}
%----------------------

%------------------------------------------------------
\begin{figure}
\begin{center}\includegraphics[width=0.4\textwidth]{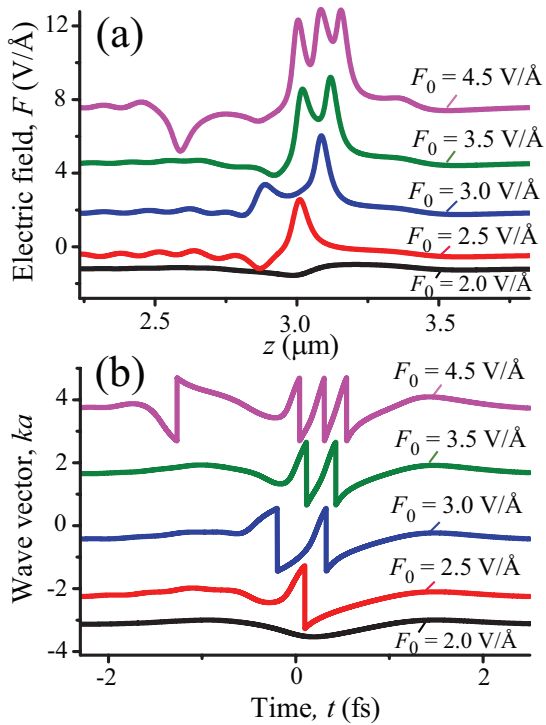}\end{center}
%\vspace*{-1cm}
\caption{Bloch oscillations in the transmitted field and electron momentum.
(a) Electric field distribution in space of the transmitted optical pulse is shown for different 
values of $F_0$. (b) The corresponding dimensionless time-dependent wave vectors in the first Brillouin zone  $ak_T(q=0,t)$ as functions of time $t$.
The origin of time is chosen arbitrary, and the graphs are offset vertically for clarity. 
}
\label{Fig_transmitted_field_wavevector_different_F}
\end{figure}
%----------------------

This enhanced transmission of the ultrastrong optical pulse is accompanied by an increase of the electric field $F^\mathrm{metal}$ inside the metal nanofilm. In Fig.\ \ref{Fig_field_inside_outside}, we show the time evolution of this field at the midplane of the nanofilm in comparison to the incident pulse. At a near-critical  pulse field $F_0=2~\mathrm{V/\AA}\sim F_c$, see Fig.\  \ref{Fig_field_inside_outside} (a), the electric field $F^\mathrm{metal}$ inside the metal is much weaker than that of the incident pulse. In contrast, for larger pulse amplitudes ($F_0=2.5, 3.5~\mathrm{V/\AA}$), see Fig.\   \ref{Fig_field_inside_outside} (b, c), the electric field $F^{\mathrm{metal}}$ becomes comparable to the incident-pulse electric field. The sharp peaks and high-frequency 
oscillations of electric field $F^{\mathrm{metal}}$ are due to the Bloch oscillations 
in electron dynamics, as we discuss below. 

The reflectance of the optical pulse (a fraction of the reflected pulse energy) is shown in Fig.\ \ref{Fig_reflectance}(a) as a function of $F_0$. Strong suppression of the pulse reflectance and correspondingly increase of the pulse transmission for $F_0 > F_c\sim 2$ V/\protect{\AA} is clearly visible. The suppression of the reflectance is correlated with increase of the electric field inside nanofilm shown by the red line in Fig.\ \ref{Fig_reflectance}(a) as $F^{\mathrm{metal}}_{\mathrm{max}}/F_0$.
The strong changes in both reflectance and internal electric field occur at  $F_0 \sim F_c\sim 2$ V/\protect{\AA}.

The absorbance of the metal nanofilm, calculated as a fraction of the pulse energy dissipated inside the nanofilm, is illustrated in Fig.\ \ref{Fig_reflectance} (b) as a function of amplitude $F_0$. This predicted behavior is very unusual. At a low pulse amplitude, the absorbance is understandably low due to the skin effect since most of the pulse is reflected back. Then as $F_0$ increases,  the absorbance increases dramatically reaching $\approx 4\%$ maximum for $F_0\approx 3~\mathrm{V/\AA}\gtrsim F_c$, which is attributed to the Bloch oscillations and WS localization leading to the collapse of the skin effect -- cf.\ the red line in panel (a). With the further increase of $F_0$, the absorbance decreases despite the field in the metal being almost the same. This can be understood as the nanofilm acquiring properties of semimetal with a low active conductivity and, consequently, low optical loss characteristic of semimetals in contrast to good (plasmonic) metals.

The power density dissipated from a single pulse in the metal reaches its maximum also at $F_0\approx 3~\mathrm{V/\AA}$. After its dissipation and thermal equilibration, this causes an estimated increase of the nanofilm temperature by $\sim 500~\mathrm{K}$. Thus the metal may survive such a high-intensity pulse without a damage. This is attributed to loss of the metallic and onset of semimetallic properties by the nanofilm. Our theory explains the absence of optical damage of metal electrodes subjected to comparable pulses in experiments \cite{Schiffrin_at_al_Nature_2012_Current_in_Dielectric}. A resistance of dielectrics to high power intensities has also been proven experimentally \cite{Schiffrin_at_al_Nature_2012_Current_in_Dielectric}. 

The origin of this highly nonlinear behavior of a metal film in a strong optical field can be understood from electron dynamics within a single conduction band.
% In the absence of the optical field, the conduction band below the Fermi surface is  occupied by electrons with wave vectors $|q|\leq  k_F$. 
In the optical field, an electron with initial wave vector $q$ is moving in the reciprocal space following the time-dependent wave vector $k_T(q,t)$, see Eq.\ (\ref{kT}) in Appendix.
Therefore,  all electrons are shifted in the reciprocal space by the same wave vector $ \Delta q (t) = \frac{e}{\hbar} \int^t  F_x(z,t_1) dt_1$, and the net current is generated. For a strong pulse, the shift $\Delta q$ is large and, for $F_0\gtrsim F_c$, can become greater than the Brillouin zone extension $k=2\pi /a$, causing the Bragg reflection of the electrons, which results in the Bloch oscillations and in the onset of the WS localization states.

At $F_0\gtrsim F_c$, the electron current acquires oscillations at Bloch frequency $\omega_B\gtrsim 4\omega$, which suppresses the susceptibility at the optical frequency $\omega$. This results in the loss of the metallic optical properties. Since the electron spectrum is discreet (the WS ladder) and, consequently, the density of states at the Fermi level is zero, the metal in the strong optical fields behaves as a semimetal with a relatively high transparency and low reflection.

%the electrons in the WS states are spatially localized with localization length $L_{WS}$ determined by the strength of the electric field, $L_{WS} \sim \frac{\hbar^2}{m a^2 eF}$. At critical field, $F_t \sim 2 $ $V/\mathrm{\protect{\AA}}$ the WS localization length is $L_{WS} \sim 2$ $\mathrm{\protect{\AA}}< a = 4$ $\mathrm{\protect{\AA}}$. The Wannier-Stark states are centered at the lattice sites. In time-varied electric field, the electrical current is generated through the dipole coupling of the nearest-neighbor WS states. In strong electric field $F> F_{cr}$ the overlap of such states is small, which results in generation of relatively weak electric current and correspondingly small reflection of the incident pulse. 

In Fig.\ \ref{Fig_transmitted_field_wavevector_different_F}(a), the spatial distribution of electric field in the transmitted optical pulse is shown for different values amplitude $F_0$.  With increasing $F_0$ above the threshold 
%$F_{2c}\approx 3~\mathrm{V/\AA}$
$F_{c}\approx 2~\mathrm{V/\AA}$, well pronounced Bloch oscillations develop in the field distribution. Their total number is proportional to field amplitude, $n\approx |e|aF_0/(2\hbar\omega)$. This is the number of times that an accelerated electron crosses the Brillouin zone boundary, as can be illustrated by the comparison to the temporal dependence of the electron quasi-momentum displayed in  Fig.\ \ref{Fig_transmitted_field_wavevector_different_F}(b). These Bloch oscillations are also visible in the temporal evolution of electric field inside the metal film --
Fig.\ \ref{Fig_field_inside_outside}. 

%In conclusion, the metal nanofilm shows quite different behavior at weak and strong intensities of the incident ultrashort optical pulse. At weak intensity, the nanofilm behaves as a regular metal with small penetration depth of an external electric field and strong reflection of the incident pulse. At strong intensity, the metal system becomes highly nonlinear with strong suppression of the generated electric currents. As a results the metallic nanofilm becomes highly transparent and behaves as a semiconductor. The critical intensity of the incident pulse at which the strong field behavior emerges is determined by the condition of formation of WS states or condition that the electrons in the CB of a metal experience the first Bragg reflection. The Bragg scattering also introduces high frequency Bloch oscillations in the transmitted optical pulse. 

To briefly conclude, we have predicted a highly unusual and interesting behavior of metal nanofilms subjected to strong ultrashort (single-oscillation) optical pulses with the  field amplitude $\sim 3~\mathrm{V/\AA}$ (intensity $\sim 2.4\times 10^{14}~\mathrm{W/cm^2}$). This includes such effects as reduction of the metallic high reflection (suppression of the skin effect) and great increase of transmission of the pulse energy through the nanofilm, while both the absorbance and the total energy deposition dramatically decrease at the high pulse intensity. This indicates that the optical field induced a transition to a semimetallic state. These phenomena develop at subcycle times $\lesssim 1~\mathrm{fs}$ and are driven by the pulse instantaneous amplitude. The metal almost returns to its original state by the end of the pulse. The transformation of the metal to a semimetal predicted in this Letter follows the optical field and is reversible. The transmitted and reflected pulses possess non-zero areas, which will cause net current (charge transfer) in media they affect. In fact, the highly-nonlinear reflection and transmission phenomena described in this Letter can be used to generate ultrashort pulses where the electric field is predominantly in one direction.  The modulation depth for the transmitted pulses is very high, which shows prospects of using metals as active elements in ultrafast modulators and field-effect transistors with petahertz bandwidth.

%%%%%%%%%%%%%%%%%%%%%

This work was supported by Grant No. DEFG02-01ER15213 from the Chemical Sciences,
Biosciences and Geosciences Division and by Grant No. DE-FG02-11ER46789 from the Materials Sciences and Engineering Division of the Office of the Basic Energy Sciences, Office of Science, U.S. Department of Energy.

\section* {Appendix}

In the presence of the optical field, $F_x (z,t)$, we 
express the general solution of the time-dependent Schr\"odinger equation in the 
basis of the Bloch functions as
\begin{equation}
\Psi(x,z,t) = \sqrt{\frac{a}{2\pi}} \sum_{\alpha = v,c}
 \int_{-\pi/a}^{\pi/a} dk ~ \phi_{\alpha }(k,z,t)
\psi_{\alpha k} (x)~, 
\label{psi0}
\end{equation}
where $\phi_{\alpha }(k,z,t)$ can be expressed in term of the Houston functions \cite{Houston_PR_1940_Electron_Acceleration_in_Lattice} $\Phi^{(H)}_{\alpha q }(k,z,t)$,
\begin{eqnarray}
&& \phi_{\alpha }(k,z,t)=\sum_q \hat{\beta} _{\alpha} (q,z,t)\Phi^{(H)}_{\alpha q }(k,z,t)~,
\label{phi_model2}\\
&  & \Phi^{(H)}_{\alpha q }(k,z,t) =
 \tilde\delta \left( k - k_T(q,t) 
% + \frac{ea}{\hbar} \int F_x(z,t) dt  - q 
                  \right)      \times    \nonumber \\
& & %\times
 \exp \left\{- i \left(  t \frac{\epsilon_{\alpha  }}{\hbar } 
 + \frac{\Delta _{\alpha }}{2 \hbar } 
 \!\! \int ^t_{-\infty} \!\! dt_1
\cos \left[ k_T(q,t_1) a \right] 
%\left( q + \frac{ea}{\hbar} \int^{t_1} \!\!\!\!  F_x(z,t_2) dt_2 \right) 
\right) \right\}.
\label{phi_h}
\end{eqnarray}
Here the time-dependent wave vector is defined as
\begin{equation} 
k_T(q,t) = q +  \frac{e}{\hbar} \int^t_{-\infty}  F_x(z,t_1) dt_1 ,
\label{kT}
\end{equation}
and $\tilde\delta  (k) = 
\sum _n \delta (k + 2\pi n/a)$, where $n=0,\pm1,\dots$, and $\delta(k) $ is the 
Dirac delta-function. 
The Houston functions are exact solutions of the time-dependent Schr\"odinger equation for a single band with the Bloch function $\psi_{\alpha q} (x)$ as the initial condition at $t=-\infty$.
% Note that the exponential phase factor in Eq.\ (\ref{phi_h}) is an adiabatic phase due to time dependence of the electron energy, $\exp ( - \frac{i}{\hbar} \int E_{\alpha } (k_T(q,t)) dt ) $.

Expansion coefficients $\hat{\beta} _{\alpha } (q,z,t)$ satisfy the equations 
\begin{equation}
\frac{d\hat{\beta} _{\alpha } (q,z,t) }{dt} =   - i\frac{F_x(z,t)}{\hbar}
\sum_{\alpha ^{\prime } \neq \alpha } 
%Z_{\alpha, \alpha ^{\prime }} 
Q_{\alpha \alpha ^{\prime }} (q,z,t) 
\hat{\beta} _{\alpha ^{\prime }} (q,z,t) 
\label{final1} ~,
\nonumber
\end{equation}
where we  denoted
\begin{eqnarray}
&& Q_{\alpha \alpha ^{\prime }} (q, z,t )  = Z_{\alpha \alpha ^{\prime }}
\exp \left\{ i\left[ t \frac{\epsilon_{\alpha  } - \epsilon_{\alpha ^{\prime } }}{\hbar} +
  \right. \right.   \nonumber \\
 && \left. \left. 
\frac{\Delta _{\alpha } - \Delta_{\alpha ^{\prime } }}{2 \hbar }  \!\! \int ^t _{-\infty}\!\! dt_1
\cos \left( q + \frac{ea}{\hbar} \int^{t_1}_{-\infty}   F_x(z,t_2) dt_2 \right)
    \right]   \right\}~,
\nonumber\\
    \label{Qq}
&&Z_{\alpha\alpha ^{\prime}} = \frac{e}{a}
\int_{-a}^{a} dz ~ u_{\alpha k}(z) ^{*} i
\frac{\partial}{\partial k} u_{\alpha ^{\prime} k}(z)~.
\end{eqnarray}
Here dipole matrix elements $Z_{\alpha\alpha ^{\prime}}$ describe diabatic coupling of VB and CB in optical field. 

The electric current generated by electron dynamics in the optical-pulse electric field  has two contributions, interband and intraband, and is expressed as
\begin{equation} 
J_x = 
J_x ^{\mathrm{inter}} +J_x ^{\mathrm{intra}}~.
\end{equation}
The interband current is 
\begin{equation}
J_x ^{\mathrm{inter}} (z,t) = \partial P_x ^{\mathrm{inter}} (z,t) /
\partial t~,
\end{equation}
where $\partial P_x ^{\mathrm{inter}} (z,t)$ is the interband polarization, which has the following form 
\begin{eqnarray}
&& P_x^{\mathrm{inter}} (z,t)= \frac{1}{2\pi a^3} 
\times  \nonumber \\
&& \int_{-\pi}^{\pi } 
dq  \sum _{\mu = v,c} f_{\mu } (q) \left[ {\cal B}^{(\mu )\dagger } (q,z,t) \hat{Q} (q,z,t) {\cal B}^{(\mu )}(q,z,t)
\right] ~,
\nonumber 
\\
\label{P0}
\end{eqnarray}
where  $\hat{Q}$ is matrix  with elements (\ref{Qq}) $Q_{\alpha \alpha ^{\prime }}$, and
 ${\cal B}^{(\mu )} = (\hat{\beta} _{v },\hat{\beta} _{c })$ is a
two-component vector, which is determined by the solution of Eq.\ (\ref{final1}) with the following 
initial conditions: ${\cal B}^{(v )} = (1,0)$ and ${\cal B}^{(c )} = (0,1)$. Here $f_{\mu } (q) $ is the 
Fermi factor, which is  1 for initially occupied states, i.e. $f_{\mu =v } (q) =1$ and $f_{\mu =c } (|q|<k_F) =1$, 
where $k_F$ is the Fermi wave vector, and it is zero otherwise. 
The intraband  current is due to shifting of electrons in space
and is expressed as
\begin{eqnarray}
&& J_x ^{\mathrm{intra}} (z,t)= \frac{1}{2\pi a^3} 
\int_{-\pi}^{\pi } dq  
 \sin \left[ k_T(q,t) a \right] \times  \nonumber \\
&& 
\sum _{\mu = v,c} f_{\mu } (q)  
 \left[ \sum _{\alpha = v,c}
 {\cal B}^{(\mu )\dagger }_{\alpha } (q,z,t)  
       \frac{\Delta_{\alpha}}{2\hbar }
  {\cal B}^{(\mu )}_{\alpha }(q,z,t)
  \right] 
~.
\nonumber
\\ \label{J0}
\end{eqnarray}

%\bibliography{references}

\end{document}